\documentclass[preprint,10pt,numbers,sort&compress]{elsarticle}
\usepackage{amsmath,amssymb,amsfonts}
\usepackage{xspace}
\usepackage{listings}
\usepackage{xcolor}
\usepackage{url,hyperref}
\usepackage{bbold}
\usepackage{slashed}
\usepackage{lmodern}

\journal{CPC}
\date{DO-TH 20/16}

\newcommand{\ARGES}{\texttt{ARGES}\xspace}
\newcommand{\SARAH}{\texttt{SARAH}\xspace}
\newcommand{\pyrate}{\texttt{PyR@TE}\xspace}
\newcommand{\mathematica}{\textit{Mathematica}\xspace}
\newcommand{\code}[1]{{\texttt{#1}}}
\newcommand{\eq}[1]{\eqref{eq:#1}}

\newcommand{\Sec}[1]{Sec.~\ref{sec:#1}}
\newcommand{\hc}{\text{h.c.}}
\newcommand{\Tr}{\mathrm{tr}}
\newcommand{\tr}{\mathrm{tr}}

\newcommand{\Nf}{N_f}
\newcommand{\Nc}{N_c}
\newcommand{\SU}[1]{SU\!\left(#1\right)}

\definecolor{LightGray}{gray}{0.93}
\definecolor{CodeGray}{gray}{0.5}
\definecolor{CodeOutput}{rgb}{0,.5,1}

\renewcommand{\paragraph}[1]{\par\vspace{0.5em}\textbf{#1}}

\begin{document}

  \begin{frontmatter}
  \title{ARGES -- Advanced Renormalisation Group Equation Simplifier}
 \author[add1]{Daniel F. Litim}
   \ead{D.Litim@sussex.ac.uk}
 \author[add2,add1]{Tom Steudtner}
  \ead{tom2.steudtner@tu-dortmund.de}
    \address[add1]{Department of Physics and Astronomy, University of Sussex, Brighton, BN1 9QH, U.K}
   \address[add2]{Fakult\"at Physik, TU Dortmund, Otto-Hahn-Str.4, D-44221 Dortmund, Germany}

  \begin{abstract}
    We present the initial release of \ARGES, a toolkit for obtaining  renormalisation group equations in perturbation theory. 
     As such, \ARGES can handle any perturbatively renormalisable four-dimensional quantum field theory.
     Notable further features include a symbolic rather than numeric computation, input of unconventional scalar and Yukawa sectors, an interactive evaluation and disentanglement as well as capabilities to inject algebraic simplification rules.   
     We provide a conceptual and practical introduction into \ARGES, and highlight similarities and differences with complementary packages. 
  \end{abstract}
  
  \end{frontmatter}

\noindent  
{\bf PROGRAM SUMMARY}

\begin{small}
\noindent
{\em Program Title:}  ARGES \\
{\em Licensing provisions:} GPLv3  \\
{\em Programming language:} Wolfram Language (Mathematica)   \\
{\em Nature of problem:} Computation of renormalisation group equations of perturbatively renormalisable quantum field theories in four space-time dimensions to various  orders in the loop expansion. \\
{\em Solution method:} Automatize computation of RGEs from input of field content, representations, couplings and index contractions.\\
{\em Unusual features:} Allows for variable generations, symmetry representations, and gauge groups. Advanced algebraic structures can be defined via simplification rules.\\

\end{small}

\section{Overview}
  In ancient greek mythology, Arges is the name of a cyclops who, after being exiled by the Titans was eventually freed by Zeus to forge lightning bolts for his battle against their former masters.
  In this tradition of being a helping hand in times of struggle, we introduce the computational tool \ARGES\ --- Advanced Renormalisation Group Equation Simplifier. 

As the name suggests, \ARGES is a framework to obtain perturbative  renormalisation group equations for renormalisable, four-dimensional quantum field theories. The task  is achieved algebraically, by extracting expressions from known formal results  in the $\overline{\text{MS}}$  renormalisation scheme  \cite{Machacek:1983tz,Machacek:1983fi,Machacek:1984zw,Luo:2002ti,Curtright:1979mg, Jack:1990eb,Jack:2018oec, Pickering:2001aq, Sperling:2013eva, Sperling:2013xqa, Mihaila:2012pz, Mihaila:2014caa, Schienbein:2018fsw, Poole:2019kcm,Steudtner:2020tzo, Sartore:2020pkk}.
 
Written in the \textit{Wolfram Language}, \ARGES has been tested for \mathematica versions 8.0 -- 12.0 \cite{Mathematica} and is available via \cite{ARGES:link} under \textit{GNU General Public License} \cite{GPLv3}. 
In this work, we explain and provide examples for \ARGES key features and capabilities, as well as highlight similarities and differences  with alternative software packages 
\SARAH 4
\cite{Staub:2008uz,Staub:2009bi,Staub:2010jh,Staub:2012pb,Staub:2013tta}
and \pyrate 3 \cite{Lyonnet:2013dna,Lyonnet:2016xiz,Sartore:2020gou}.

\section{Statement of the problem  \& conventions} 
The computation of renormalisation group equations (RGEs) of running couplings continues to be of key importance for many applications of quantum field theory (QFT) in particle physics and critical phenomena. 
Thankfully, for weakly coupled and perturbatively renormalisable QFTs, the procedure can be partially automatised. This is achieved by considering a template action in which any concrete model can be embedded,  
which allows the computation to be broken down into two parts. In a first step, momentum integrals and spinor traces are resolved, and all RGEs expressed in terms of generalised couplings. Universal expressions of this type have been made available in the literature \cite{Machacek:1983tz,Machacek:1983fi,Machacek:1984zw,Luo:2002ti,Curtright:1979mg, Jack:1990eb,Jack:2018oec, Pickering:2001aq, Sperling:2013eva, Sperling:2013xqa, Mihaila:2012pz, Mihaila:2014caa, Schienbein:2018fsw, Poole:2019kcm,Steudtner:2020tzo, Sartore:2020pkk}, and can be mapped back onto any renormalisable QFT in a second step. 

In practice, however, the embeddings of couplings and fields are in general non-trivial.  In fact, for increasingly complex  QFTs involving a large number of quantum fields and interaction terms, the second step becomes rapidly more involved than the first one. Moreover, with increasing loop order, the complexity increases even further. 
For these reasons, computer codes such as \ARGES are required, which automatise the embedding and its resolution to enable universal access to the results in the literature and the extraction of RGEs for concrete models.

With this aim in mind, \ARGES 
adopts the conventions developed in \cite{Machacek:1983tz,Machacek:1983fi,Machacek:1984zw} and expresses the template action for a perturbatively renormalisable  four-dimensional QFT as
  \begin{equation}\label{eq:master-template}
  \begin{aligned}
    \mathcal{L} =
     & -\frac{1}{4} F_I^{\mu\nu} F^I_{\mu \nu} + \frac{1}{2} D^\mu \phi_a D_\mu \phi_a + i \psi_j^\dagger \sigma^\mu D_\mu \psi_j + \mathcal{L}_\text{gh} + \mathcal{L}_\text{gf}\\
     & - \frac{1}{2} \left(Y^a_{jk} \,\psi_j \varepsilon \psi_k \phi_a + Y^{a*}_{jk} \,\psi_j^\dagger \varepsilon \psi_k^\dagger \phi_a\right) - \frac{1}{4!} \lambda_{abcd} \,\phi_a \phi_b \phi_c \phi_d\,\\
     &  -\frac{1}{2} \left[m_{jk}\, \psi_j \varepsilon \psi_k + m^*_{jk} \,\psi^\dagger_j \varepsilon \psi^\dagger_k \right]
-\frac{\mu^2_{ab}}{2!} \phi_a \phi_b - \frac{h_{abc}}{3!} \phi_a \phi_b \phi_c\,.
  \end{aligned}
  \end{equation}
  Here $\mathcal{L}_\text{gf, gh}$ are gauge fixing and ghost terms and $\psi_i$, $\phi_a$ being Weyl fermions and real scalar fields.
  Their respective universal indices $i,\,j,\,k\cdots$ and $a,\,b,\,c\cdots$ run over all particle species (including real and imaginary parts of complex scalars), generations, flavours, and gauge components.\footnote{An interesting and slightly modified  basis has recently been put forward in \cite{Poole:2019txl,Poole:2019kcm}. However, the development of \ARGES predates these works.} In the same spirit, a sum over all gauge groups is implied. Gauge interactions are given by
  \begin{equation}
  \begin{aligned}
    D_\mu\phi_a &= \partial_\mu \phi_a + i A_\mu^I \, t^I_{ab} \, \phi_b, \\
    D_\mu\psi_i &= \partial_\mu \psi_i + i A_\mu^I \, t^I_{ij} \, \psi_j, \\
    F_{\mu\nu}^I &= \partial_\mu A^I_\nu - \partial_\nu A^I_\mu -  t^{I}_{\,JK} \,A^J_\mu A^K_\nu,
  \end{aligned}
  \end{equation}
  where $t^I$ are generators of the respective representations, and gauge couplings $g_\alpha$ are implied via $A_\mu^I = g_\alpha A_\mu^{I_\alpha}$, with $\alpha$ counting over all gauge subgroups and indices $I,J,K \cdots $ denote the adjoint indices of gauge fields. Finally  $\mu,\,\nu \cdots $ are the usual Lorentz indices and $\varepsilon = \left(\begin{smallmatrix}0 & 1\\ -1 & 0\end{smallmatrix}\right)$ denotes the two-dimensional Levi-Civita symbol which contracts the implicit spinor components of fermions $\psi_i$.
Overall, any allowed interaction monomial $\sim \alpha_{\ell_1 \cdots  \ell_n} X_{\ell_1} \cdots  X_{\ell_n}$ involving scalar, fermion and vector fields $X_\ell \in \{\phi_a,\,\psi_i,\, A_\mu^I \}$ as well as couplings  $\alpha_{\ell_1 \cdots  \ell_n} \in \{ g_\alpha,\,Y^a_{jk},\,\lambda_{abcd},\,m_{jk},\,\mu^2_{ab},\,h_{abc}\}$ (with the appropriate set of indices $\ell_i$) can be accommodated. 
  After renormalisation, these bare couplings are replaced by the respective renormalised ones $\alpha_{\ell_1 \cdots  \ell_n} (Q)$  and field strength renormalisation constants $\sqrt{Z}_{\ell_1 \ell_2}$ are introduced
  \begin{equation}
    X_{\ell_1} \mapsto \left(\sqrt{Z}_{\ell_1 \ell_2}(Q) + \delta \sqrt{Z}_{\ell_1 \ell_2} \right)X_{\ell_2}, \quad  \alpha_{\ell_1 \cdots  \ell_n} \mapsto \alpha_{\ell_1 \cdots  \ell_n}(Q) + \delta \alpha_{\ell_1 \cdots  \ell_n}.
  \end{equation}
  Here, $\delta \sqrt{Z}$ and $\delta \alpha$ are counter-terms, and $Q$ denotes the renormalisation scale parameter.
  The renormalisation scale running of $\sqrt{Z}(Q)$ and $\alpha(Q)$ are then encoded in the respective field anomalous dimensions and $\beta$-functions
  \begin{equation}\label{eq:RGE}
    \gamma_{\ell_1 \ell_2}(\alpha) = \left(\frac{d \ln \sqrt{Z}(Q)}{d \ln Q}\right)_{\ell_1 \ell_2} , \qquad \beta_{\ell_1 \cdots  \ell_n} (\alpha) = \left(\frac{d \,\alpha(Q)}{d \ln Q}\right)_{\ell_1 \cdots  \ell_n}.
  \end{equation} 
Presently,  \ARGES implements the literature expressions for \eq{RGE} up to three-loop order in gauge coupling $\beta$-functions \cite{Machacek:1983tz,Luo:2002ti,Curtright:1979mg, Pickering:2001aq, Mihaila:2012pz, Mihaila:2014caa, Poole:2019kcm}. RG equations for Yukawa interactions \cite{Machacek:1983fi,Luo:2002ti,Schienbein:2018fsw}, scalar quartics \cite{Machacek:1984zw,Luo:2002ti,Schienbein:2018fsw,Steudtner:2020tzo} and cubics \cite{Machacek:1984zw,Luo:2002ti,Schienbein:2018fsw,Steudtner:2020tzo}, fermion and scalar masses \cite{Machacek:1983fi,Machacek:1984zw,Luo:2002ti,Schienbein:2018fsw,Steudtner:2020tzo}, vacuum expectation values \cite{Sperling:2013eva, Sperling:2013xqa} as well as scalar and fermion anomalous dimensions \cite{Machacek:1983fi,Machacek:1984zw,Luo:2002ti,Steudtner:2020tzo} are implemented up to two-loop order. Finally, purely scalar contributions to scalar quartic and cubic interactions, masses, and field anomalous dimensions are included up to four-loop order \cite{Jack:1990eb,Jack:2018oec,Steudtner:2020tzo}. 
All of the above expressions have originally been obtained using dimensional regularisation \cite{Bollini:1972bi,Bollini:1972ui} and the $\overline{\text{MS}}$ renormalisation scheme \cite{tHooft:1973mfk,Bardeen:1978yd}.   
  
A central capability of \ARGES is  to extract RGEs
for the unembedded couplings $\alpha^{(i)}$. In general, these can be \textit{disentangled} from the linear combination of generalised $\beta$-functions \eq{RGE} by determining the coefficients $\eta^{(i)}_{\ell_1 \cdots  \ell_n}$ of the couplings 
  \begin{equation}\label{eq:disentangle-tree}
    \alpha^{(i)} = \sum_{\ell_1 \cdots  \ell_n} \eta^{(i)}_{\ell_1 \cdots  \ell_n} \alpha_{\ell_1 \cdots  \ell_n} 
  \end{equation}
at tree level and then employing them for the loop-level RGEs
  \begin{equation}\label{eq:disentangle-loop}
  \beta^{(i)}(\alpha^{(j)}) \equiv \frac{d\,\alpha^{(i)}}{d \ln Q} = \sum_{\ell_1 \cdots  \ell_n} \eta^{(i)}_{\ell_1 \cdots  \ell_n} \beta_{\ell_1 \cdots  \ell_n} .
  \end{equation}
  In the following, implementation details will be highlighted in comparison with pre-existing software.  
  \section{Design \& comparison with other frameworks}
  The design goals of \ARGES are complementary to those of the two alternative software packages available with the same scope, namely \SARAH and \pyrate. 
Broadly speaking,  \SARAH and \pyrate are quite different in their implementation: \pyrate is a highly specialised \textit{python} package accepting a single input file and command line parameters. It allows for theories ranging from simplest toy models to complicated SM extensions.
 \SARAH on the other hand is a large \mathematica package and in fact a wider framework with many more capabilities. Geared towards realistic theories, it requires complex inputs, and is the back end of various farther reaching scientific software.
On the other hand, the inner workings and capabilities of \pyrate and \SARAH are conceptually similar. It is here where \ARGES takes a  different stance. The key differences in \ARGES functionality will be highlighted in the following.
 
 \paragraph{Index contractions are user input.} With the exception of interaction terms involving gauge bosons,  contractions  of gauge and global indices over each interaction vertex have to be specified as input. 
\ARGES knows only very little about the Lie algebras, and shifts the responsibility of formulating a gauge invariant action, consistent with the desired symmetries, to the user. 
 The advantage is that the user has full control over the shape of the action, eliminating any uncertainty e.g.~about coupling normalisations. 
 While \SARAH and \pyrate allow for an explicit input of contractions, they also link to the packages \code{Susyno} \cite{Fonseca:2011sy} and its \textit{python} clone \code{PyLie} \cite{pylie:link,Sartore:2020gou} that automatise the search for index contractions. However, this procedure provides a much less fine-grained control, especially in cases where several contractions are possible.
  Even though the automatisation avoids explicit violations of gauge invariance, all three codes may give inconsistent results if not all interactions allowed by the symmetries are manually included from the outset.
 
   \paragraph{Groups, representation and multiplicities can be variables.} As opposed to \pyrate and \SARAH, \ARGES allows for gauge groups to be either completely undetermined, formulating the results in terms of general gauge invariants, or to be an entire family of gauge groups such as $SU(N)$ without specifying $N$. More so, the representations for each field can be kept a variable as well, and so does the number of generations or matter field multiplicities. This ties in with the previous point: \ARGES is agnostic to symmetries and algebras, and treats everything as an index. This allows for a systematic study of a large range of models and represents a key advantage over existing codes.

  \paragraph{Gauge invariants are not resolved by default.} \ARGES does not resolve gauge invariants such as Dynkin indices and quadratic Casimirs by default. In general, they are also considered user input. For simple representations of $U(1)$, $SU(N)$, $SO(N)$ and $Sp(2N)$ gauge groups, invariants can be computed automatically and are available as substitution rules. 
 
 \paragraph{Scalars can carry flavour.} In \SARAH and \pyrate, only fermions carry a single flavour index, while scalars do not. In \ARGES, scalars carry two such indices instead, which can be contracted freely, and whose ranges can be kept as open parameters. This is sufficient to embed any scalar field allowed by \eq{master-template}, including with parametrical global multiplicities.
 
 \paragraph{Flexible flavour structure in Yukawa sector.} While the design of \SARAH and \pyrate constrains Yukawa couplings to be of matrix form $y_{ij} \,\overline{\psi}_i\,\phi\,\psi'_j  + \hc$~in the fermion generations, \ARGES also allows for a generalised contraction $h$ of flavour indices and a single Yukawa coupling $y$ in a vertex $ y \,h_{ijkl} \,\overline{\psi}_i\,\phi_{kl}\,\psi'_j  + \hc $ 
 This ensures that \ARGES' capabilities completely cover the set of actions \eq{master-template}, 
including \textit{e.g.}~BSM models with extended flavour structures such as in GUTs, or in settings where the multiplicities of  gauge or matter fields remain free parameters throughout. The latter features are  absent in \SARAH and \pyrate.\footnote{Finite tensorial index structures can be flattened-out   by expanding  flavour contractions at each vertex, and by treating flavour components of couplings and fields as independent vertices and particles to make this accessible for  \SARAH. In \pyrate, this can be done if fermion generations are integers and not open parameters.}
 
  \paragraph{Disentanglement of RGEs by the user.} In \pyrate and \SARAH, the disentanglement of couplings and RGEs as described in \eq{disentangle-tree} and \eq{disentangle-loop} is fully automatised. 
  However, such algorithms can be computationally expensive for even unsuccessful. This may occur in valid models, but is bound to happen if the specified couplings cannot be disentangled.\footnote{The latest version of \pyrate offers a very reliable and well-performing algorithm  with couplings being flagged if they  cannot be disentangled.} 
  In \ARGES, the disentangling of couplings and RGEs is out-sourced to the user (examples are provided  in \Sec{Example}).
  Besides enhancing the stability of the code, this strategy bears two further advantages: for one, there are often several choices of external indices, and the user is free to choose the one minimising the computational efforts for the RGEs, e.g.~with the most symmetries of external legs. Secondly, the setup allows the user to input external indices as variables, and reconcile tree- with loop-level contractions. This also offers a simple way to check whether couplings allowed by the symmetry but omitted in the action are nevertheless switched on by quantum fluctuations under the renormalisation group.

  \paragraph{Efficient handling of unspecified interactions.} \ARGES allows index contraction to be merely defined by a number of algebraic relations, providing a mechanism to insert such information and allow for an efficient computation of RGEs (an example for this is provided  in \Sec{Example}).
 
  \paragraph{Irreducible representations, kinetic mixing and gauge fixing.} At the time of writing, the capabilities of \ARGES are limited to irreducible representations without kinetic mixing of $U(1)$ gauge groups, which is implemented in the most recent versions of  \SARAH and \pyrate.  All three packages assume an $R_\xi$ gauge fixing. 

  \section{Installation}
  \ARGES is designed to be easily distributable, the code has no external dependencies and is located in a single file \texttt{ARGES.m}. Moreover, another design goal is that \ARGES does neither require nor encourage a notebook, or any graphical user interface in general, access to a \mathematica kernel is sufficient. The source code can be acquired from \cite{ARGES:link}, e.g.~by cloning the git repository.
  \begin{lstlisting}
  git clone https://github.com/TomSteu/ARGES
  \end{lstlisting}
  The relevant file can be loaded by the kernel directly via 
  \begin{lstlisting}
   Get["~/path/to/ARGES.m", ARGES`];
  \end{lstlisting}
  or alternatively, moved into a location contained in \texttt{\$Path} manually, using \texttt{Install[\_\_]} or the graphical user interface and included via:
    \begin{lstlisting}
   <<ARGES`
  \end{lstlisting}
  If no output is produced by whatever the method of choice, then the installation was successful. Next, we will proceed with an example on how to define a valid input model.
  \section{ARGES by example}\label{sec:Example}
In this section we demonstrate the basic and advanced functionality of \ARGES by application, in good faith that generalisations to other models are then largely straightforward.

  \subsection{Defining a model}
  The input required by \ARGES can either be provided by a model file or by an interactive session.  
  Here, we proceed with the model
  \begin{equation}\label{eq:LiSa-action}
  \begin{aligned}
    \mathcal{L} =
     & -\tfrac{1}{4} F_A^{\mu\nu} F^A_{\mu \nu} + \overline{Q}\,  i\slashed{D}\, Q +  \Tr\left[\partial_\mu \phi ^\dagger  \partial^\mu \phi\right]  + \mathcal{L}_\text{gh} + \mathcal{L}_\text{gf}\\
     & - y \left[ Q_i^{L\dagger} \phi_{ij} Q_j^R + Q_i^{R\dagger} \phi^\dagger_{ij} Q^L_{j} \right]- u \,\tr\left[\phi^\dagger \phi \phi^\dagger \phi\right] - v \,\tr\left[\phi^\dagger \phi \right] \tr\left[\phi^\dagger \phi \right]
  \end{aligned}
  \end{equation}
  featuring a  $\SU{\Nc}$ gauge sector, $\Nf$ quark-like Dirac fermions as well as a complex and uncharged $\Nf \times \Nf$  scalar meson matrix, coupled together by a single Yukawa interaction and two scalar quartic couplings.
The family of perturbatively renormalisable models \eqref{eq:LiSa-action} contains scalar, fermionic, and gauge degrees of freedom; it is known to display regimes with asymptotic freedom, infrared freedom, and regimes with interacting IR and UV fixed points, depending on the field multiplicities $N_c$ and $N_f$. The models  \eqref{eq:LiSa-action}  have received some attention in recent years as they provide perturbatively controlled examples for asymptotically safe QFTs in four spacetime-dimensions  \cite{Litim:2014uca,Litim:2015iea,Bond:2017tbw}.\footnote{Further works cover large-$N$ equivalences \cite{Bond:2019npq},  extensions to semi-simple gauge groups \cite{Bond:2017lnq}, supersymmetry \cite{Bond:2017suy}, and SM extensions \cite{Bond:2017wut,Hiller:2019mou,Hiller:2020fbu,Bissmann:2020lge}.}

 The model \eq{LiSa-action} has  been chosen to highlight some of \ARGES key capabilities over complementary packages. Specifically, neither \SARAH nor \pyrate can handle general $\SU{\Nc}$ gauge groups with $N_c$ unspecified, and neither of them  can define a $\Nf \times \Nf$ two-index scalar field  as input. Also, $\Nf$ Dirac fermions with $\Nf$ unspecified cannot be handled by \SARAH (but \pyrate3).
Finally, specifying the correct contractions of the Yukawa vertex and quartics is problematic in both packages, and so is the disentangling of the quartic RGEs.\footnote{However, as long as only the intact $\SU{\Nf} \times \SU{\Nf}$ flavour symmetry is considered, there is a trick to promote these to two gauge symmetries with vanishing couplings which should in principle allow model input for specific values of $N_f$ and $N_c$.}

  For \ARGES, no such limitations apply which we start to illustrate by loading the code.
  \begin{lstlisting}[numbers=left]
   <<ARGES`
   Reset[];
  \end{lstlisting}
  \ARGES is stateful, and \texttt{Reset[]} wipes any previous input without affecting the kernel memory. It is not necessary to invoke it for the first run, but recommended in notebooks as cells may be re-evaluated. Next, the gauge sector will be specified, starting with the number of gauge groups.
  \begin{lstlisting}[numbers=left, firstnumber=3, escapeinside={|}{|}]
   NumberOfSubgroups = 1;
   Gauge[g, SU[Nc], {Nc^2 - 1}];
  \end{lstlisting}
  Then, the functions \texttt{Gauge[\_\_]} are called once for each gauge group, which implies the ordering of gauge indices. The first argument is the symbol of the gauge coupling, followed by the group, or any place holder for an unknown unspecified one thereof. Finally a list of length \texttt{NumberOfSubgroups} denotes the multiplicity of the gauge bosons\footnote{In the $U(1)$ case, it marks the charge instead.} under each gauge group, in the implied ordering. Next, the matter content can be specified. We will start with registering the fermions in terms of their Weyl components.
  \begin{lstlisting}[numbers=left, firstnumber=5]
   WeylFermion[QL, Nf, {Nc}];
   WeylFermion[QR, Nf, {Nc}];
  \end{lstlisting}
  Hereby, the first argument is the name of the fermion, the second the number of flavours and the third a list of its gauge multiplicities in the same order as before. For $U(1)$ gauge groups, the charge of the field is to be inserted here. Scalar matter can be inserted with 
  \begin{lstlisting}[numbers=left, firstnumber=7]
   ComplexScalar[H, {Nf, Nf}, {1}];
  \end{lstlisting}
  which will add two real components \texttt{Re[H]} and \texttt{Im[H]}. Whenever the complex field is specified somewhere, the decomposition $\texttt{H} = \left(\texttt{Re[H]} + i \texttt{Im[H]} \right)/\sqrt{2}$ is then automatically inserted.
Alternatively, one may add the components manually via \texttt{RealScalar[\_,\_,\_]} with the same calling conventions. The syntax is similar to the fermionic case, only the second argument is now a list of two elements, as scalars always carry two flavour indices. We are now in the position to add interactions. A Yukawa term with a single coupling \texttt{y} is inserted via
  \begin{lstlisting}[numbers=left, firstnumber=8]
   Yukawa[y, H, adj[QL], QR, {KroneckerDelta[#2,#3]&}, (KroneckerDelta[#1,#3] KroneckerDelta[#2,#4])& ];
  \end{lstlisting}
  The second to fourth arguments represent the scalar and fermionic fields involved. The fifth argument is a list of \texttt{NumberOfSubgroups} elements, each being a function of three arguments representing the contractions of gauge indices of the scalar and two fermionic fields involved, in the order of appearance at this vertex. To optimise the simplification, \mathematica's built-in function $\texttt{KroneckerDelta[\_,\_]}$ should be used for each contraction. The final argument is the contraction function of flavour indices and expects four arguments, the first two being indices of the scalar, and the second two for each of the fermions, again in the order of appearance.  Obviously, we have used \mathematica's capability to define anonymous functions as \texttt{(\_)\&}, with the $n$th argument denoted by $\#n$. In the same manner, quartics can now be added, keeping in mind that scalars have two flavour indices each. 
    \begin{lstlisting}[numbers=left, firstnumber=9]
   ScalarQuartic[u, adj[H], H, adj[H], H, {1&}, (KroneckerDelta[#2,#3] KroneckerDelta[#4,#5] KroneckerDelta[#6,#7] KroneckerDelta[#8,#1])& ];
   ScalarQuartic[v, adj[H], H, adj[H], H, {1&}, (KroneckerDelta[#2,#3] KroneckerDelta[#4,#1] KroneckerDelta[#6,#7] KroneckerDelta[#8,#5])& ];
  \end{lstlisting}
  The model is now defined, and we can already compute all the gauge invariants available via
      \begin{lstlisting}[numbers=left, firstnumber=11]
   ComputeInvariants[];
  \end{lstlisting}
  which will be stored as a substitution rule in \texttt{subInvariants}.
  
  In addition, the functions \texttt{ScalarCubic[\_\_]}, \texttt{ScalarMass[\_\_]} and \texttt{FermionMass[\_\_]} allow to specify couplings $h_{abc}$, $\mu^2_{ab}$ and $m_{ij}$ in \eq{master-template}. For instance, a scalar mass term
   \begin{lstlisting}[numbers=left, firstnumber=12]
   ScalarMass[m2, adj[H], H, {1&}, (KroneckerDelta[#2,#3] KroneckerDelta[#4,#1])& ];
  \end{lstlisting}
  is compatible with the global symmetry of \eq{LiSa-action}.
  
  \vspace{.5em}
  
  Alternatively, models with more conventional Yukawa and fermion mass matrices can be entered via  
  \texttt{YukawaMat[\_\_]} and \texttt{FermionMassMat[\_\_]} respectively. For instance, a gauge theory with fermions $\psi_{L,R}$ and complex singlet scalar $\phi$ can contain a Yukawa term $\tfrac1{n} Y_{ij} (\psi_L^\dagger)_{i a} \,\phi \, (\psi_R)_{j a} + \hc$ Here, generations are counted via $i,j, \cdots $ while $a$ refers to gauge components of the fermions, and $n$ is a normalisation factor. This vertex is entered as
   \begin{lstlisting}[]
   YukawaMat[Y, phi, adj[psiL], psiR, {KroneckerDelta[#2,#3]&}, n];
  \end{lstlisting}
  into \ARGES. Moreover, 
   \begin{lstlisting}[]
   VEV[v, Re[phi], {1,1,1}, 1/Sqrt[2]];
  \end{lstlisting}
  assigns a correctly normalised VEV to the real part of the scalar, while the third argument relates to the two generation indices and the gauge index of the affected component.

  \subsection{Obtaining output}
  Now, the model \eq{LiSa-action} from the previous section
  \begin{lstlisting}[numbers=left]
   <<ARGES`
   Reset[];
   NumberOfSubgroups = 1;
   Gauge[g, SU[Nc], {Nc^2 - 1}];
   WeylFermion[QL, Nf, {Nc}];
   WeylFermion[QR, Nf, {Nc}];
   ComplexScalar[H, {Nf, Nf}, {1}];
   Yukawa[y, H, adj[QL], QR, {KroneckerDelta[#2,#3]&}, (KroneckerDelta[#1,#3] KroneckerDelta[#2,#4])& ];
   ScalarQuartic[u, adj[H], H, adj[H], H, {1&}, (KroneckerDelta[#2,#3] KroneckerDelta[#4,#5] KroneckerDelta[#6,#7] KroneckerDelta[#8,#1])& ];
   ScalarQuartic[v, adj[H], H, adj[H], H, {1&}, (KroneckerDelta[#2,#3] KroneckerDelta[#4,#1] KroneckerDelta[#6,#7] KroneckerDelta[#8,#5])& ];
   ComputeInvariants[];
  \end{lstlisting}
  will be analysed up to one loop in the gauge, Yukawa and scalar couplings. The $\beta$-functions for the gauge coupling \texttt{g} is obtained via 
  \begin{lstlisting}[escapeinside={|}{|}]
|\color{CodeGray}\textbf{In[1]:=}|   (4 |{$\mathtt{\pi}$}|)^2 |{$\beta$}|[g, 1] //. subInvariants // Expand
|\color{CodeGray}\textbf{Out[1]=}|  |\color{CodeOutput} -11/3 Nc g\verb!^!3 + 2/3 Nf g\verb!^!3|
  \end{lstlisting}
  where the second argument indicates the loop order. For any other couplings, RG equations are extracted by specifying external fields and their indices. In order to properly normalise and disentangle the system of couplings, this has to be done at tree level first. For the Yukawa interaction, the syntax is
  \begin{lstlisting}[escapeinside={|}{|}]
|\color{CodeGray}\textbf{In[2]:=}|   |$\beta$|[Re[H], adj[QL], QR, {i1, i2, 1}, {j1, a}, {j2, b}, 0]
|\color{CodeGray}\textbf{Out[2]=}|   |\color{CodeOutput} y $\delta_{\texttt{i1},\texttt{j1}}$ $\delta_{\texttt{i2},\texttt{j2}}$ $\delta_{\texttt{a},\texttt{b}}$ / Sqrt[2] |
  \end{lstlisting}
  where the first three arguments specify the scalar and fermions at the vertex of interest, followed by lists of the quantum numbers of each field, in that order. The leading elements of that list are flavour (one for fermions, two for scalars), and the remaining components gauge indices.\footnote{In case of a $U(1)$ gauge group, a dummy index $1$ has to be provided.} Finally the last argument is again the loop order, with $0$ indicating tree-level. From this output, the desirable normalisation and index structure can be read off.
  
   Now, the one-loop $\beta$-function of \texttt{y} is extracted by changing the last argument, and taking the correct normalisation into account.
  \begin{lstlisting}[escapeinside={|}{|}]
|\color{CodeGray}\textbf{In[3]:=}|   (4 |{$\mathtt{\pi}$}|)^2 Sqrt[2] |$\beta$|[Re[H], adj[QL], QR, {1,1,1}, {1,1}, {1,1}, 1] //. subInvariants // Expand
|\color{CodeGray}\textbf{Out[3]=}|   |\color{CodeOutput} 3/Nc g\verb!^!2 y - 3 Nc g\verb!^!2 y + Nc y\verb!^!2 conj[y] + Nf y\verb!^!2 conj[y]|
  \end{lstlisting}
  For the scalar quartics, syntax and procedure is very similar. As there are two couplings $u$ and $v$, we need to compute two vertices at tree level and disentangle the quartics. One of many possible choices is
    \begin{lstlisting}[escapeinside={|}{|}]
|\color{CodeGray}\textbf{In[4]:=}|   |$\beta$|[adj[H], H, adj[H], H, {1,1,1}, {1,1,1}, {1,1,1}, {1,1,1}, 0] // Expand
|\color{CodeGray}\textbf{Out[4]=}|   |\color{CodeOutput}u + v|
|\color{CodeGray}\textbf{In[5]:=}|   |$\beta$|[adj[H], H, adj[H], H, {2,2,1}, {2,2,1}, {1,1,1}, {1,1,1}, 0] // Expand
|\color{CodeGray}\textbf{Out[5]=}|   |\color{CodeOutput}v/3|
  \end{lstlisting}
  which implies that one-loop expressions for the quartics can be obtained via the following expression.
  \begin{lstlisting}[escapeinside={|}{|}]
|\color{CodeGray}\textbf{In[6]:=}|   bv = 3 (4 |{$\mathtt{\pi}$}|)^2 |$\beta$|[adj[H], H, adj[H], H, {2,2,1}, {2,2,1}, {1,1,1}, {1,1,1}, 1] //. subInvariants // Expand
|\color{CodeGray}\textbf{Out[6]=}|   |\color{CodeOutput}12 u\verb!^!2 + 16 Nf u v + 16 v\verb!^!2 + 4 Nf\verb!^!2 v\verb!^!2 + 4 Nc v y conj[y]|
|\color{CodeGray}\textbf{In[7]:=}|   bu = - bv + (4 |{$\mathtt{\pi}$}|)^2 |$\beta$|[adj[H], H, adj[H], H, {1,1,1}, {2,2,1}, {1,1,1}, {1,1,1}, 1] //. subInvariants // Expand
|\color{CodeGray}\textbf{Out[7]=}|   |\color{CodeOutput}8 Nf u\verb!^!2 + 24 u v + 4 Nc u y conj[y] - 2 Nc y\verb!^!2 conj[y]\verb!^!2|
  \end{lstlisting}
  
  In fact, the syntax $\beta$\texttt{[\_\_]}, with the leading arguments being all fields at the vertex followed by their respective lists of indices, is also utilised to obtain scalar cubic interactions as well as fermion and scalar mass terms. 
  To avoid confusion with the latter, scalar and fermionic field anomalous dimensions 
    \begin{lstlisting}[escapeinside={|}{|}]
|\color{CodeGray}\textbf{In[8]:=}|   (4 |{$\mathtt{\pi}$}|)^2 |$\gamma$|[adj[H], H, {1,1,1}, {1,1,1}, 1] //. subInvariants // Expand
|\color{CodeGray}\textbf{Out[8]=}|   |\color{CodeOutput}Nc y conj[y]|
|\color{CodeGray}\textbf{In[9]:=}|   (4 |{$\mathtt{\pi}$}|)^2 |$\gamma$|[adj[QL], QL, {1,1}, {1,1}, 1] //. subInvariants // Expand
|\color{CodeGray}\textbf{Out[9]=}|   |\color{CodeOutput}- $\mathtt{\xi}$ g\verb!^!2 /(2 Nc) + Nc $\mathtt{\xi}$/2 g\verb!^!2 + Nf/2 y conj[y] |
  \end{lstlisting}
  use the syntax $\gamma$\texttt{[\_\_]} instead.
  The output obtained in this section matches the analytic results in \cite{Litim:2014uca}, and it is straightforward to extend results to higher loop orders \cite{Bond:2017tbw}.
  
  This concludes the presentation of \ARGES basic input and output functionality, which should be sufficient for a first instruction. In the next section we will turn towards a more advanced example, demonstrating a more specialised  use of \ARGES.

  \subsection{Advanced capabilities}
  Under the hood, \ARGES optimises index summations by employing simplification rules like resolving \texttt{KroneckerDelta} contractions directly instead of brute-forcing \texttt{Sum[\_\_]}. This is critical for \ARGES performance, as such contraction sums are typically very long. Internally, they are simplified as expressions of the shape \texttt{SimplifySum[\_\_]}, and only in the last step converted back to \mathematica's built-in function \texttt{Sum[\_\_]} (the syntax is compatible), if required.\footnote{Disabling this last step with \texttt{DisableNativeSums[]} may in fact give a performance boost.} This simplification can be enforced on any expression using \texttt{SimplifyProduct[\_]}.
  
  The user may inject custom simplification rules into this mechanism by adding them to the list \texttt{subSimplifySum}. This will be demonstrated by example of a theory with a $SU(n)$ global symmetry and real scalars in the adjoint. The action of interest is
  \begin{equation}\label{eq:advanced-model}
  \begin{aligned}
    \mathcal{L} = &\tfrac12 \partial_\mu \phi^A \partial^{\mu} \phi^A - \tfrac12 m^2 \phi^A \phi^A 
    \\&- \tfrac14 \lambda_1 \left(\phi^A \phi^A\right)^2 - \tfrac12 \lambda_2  \,\phi^A \phi^B \phi^C \phi^D \left(T^{ABCD} + T^{DCBA}\right),
  \end{aligned}
  \end{equation} 
  where the object
  \begin{equation}
    T^{ABC\dots} = \tr\left[ t^A t^B t^C \dots \right].
  \end{equation} 
  are traces of $t^A$, the generator  of the fundamental representation of $SU(n)$. These structures are difficult to resolve in general, but due to the Fierz relation
  \begin{equation}
    t_{ab}^A t_{cd}^A = \tfrac12 \left( \delta_{ad} \delta_{bc} - \tfrac1{n} \delta_{ab} \delta_{cd}\right),
  \end{equation}
 together with the Dynkin index definition and the tracelessness of the generators, the following relations can be inferred
 \begin{equation}\label{eq:advanced-relations}
   \begin{aligned}
     &T = \tr\left[\mathbb{1}\right] = n, \\
     &T^A = 0, \\
     &T^{AB} = \tfrac12 \delta^{AB}, \\
     &\sum_A T^{B_1\cdots B_i\, A \,C_1\cdots C_j \,A \,D_1\cdots D_k} =\\
     &\qquad  \tfrac12 \left(T^{B_1\cdots B_i \, D_1\cdots D_k} T^{C_1\cdots C_j} - \tfrac1{n} T^{B_1\cdots B_i \, C_1\cdots C_j \, D_1\cdots D_k} \right) \\
          &\sum_A T^{B_1\cdots B_i \,A \,C_1\cdots C_j} T^{ D_1\cdots D_k \,A \,E_1\cdots E_l} = \\
          &\qquad \tfrac12 \left(T^{B_1\cdots B_i \, E_1\cdots E_l \, D_1\cdots D_k \, C_1\cdots C_j}  - \tfrac1{n} T^{B_1\cdots B_i \, C_1\cdots C_j} T^{ D_1\cdots D_k \, E_1\cdots E_l} \right).
   \end{aligned}
 \end{equation} 
  This also holds true when index ranges such as $B_1\cdots B_i$ are empty. Using \ARGES, the model \eq{advanced-model} can be analysed by entering \eq{advanced-relations} into \texttt{subSimplifySum} using patterns. It can be assumed that sums over gauge and flavour indices are expanded and each term contained in \texttt{SimplifySum[\_\_]} expression which is not nested.
  \begin{lstlisting}[escapeinside={|}{|}, numbers=left ] 
    <<ARGES`
    Reset[];
    
    NumberOfSubgroups=0;
    RealScalar[phi, {n^2-1,1}, {}];
    ScalarMass[m2, phi, phi, {}, KroneckerDelta[#1,#3]/2&];
    ScalarQuartic[|$\lambda$|1, phi, phi,  phi, phi, {}, KroneckerDelta[#1,#3] KroneckerDelta[#5,#7]/4&];
    ScalarQuartic[|$\lambda$|2, phi, phi,  phi, phi, {}, (T[#1,#3,#5,#7] + T[#7,#5,#3,#1])/2&];
   
    subSimplifySum = {
       
       T[] -> n,
       T[_] :> 0,
       T[A_, B_] :> 1/2 KroneckerDelta[A,B],
       
       SimplifySum[c_ T[B___, A_, C___, A_, D___], sum1___, {A_, __}, sum2___] :> SimplifySum[c/2 (T[B, D] T[C] - T[B, C, D]/n ), sum1, sum2],
       SimplifySum[T[B___, A_, C___, A_, D___], sum1___, {A_, __}, sum2___] :> SimplifySum[1/2 (T[B, D] T[C] - T[B, C, D]/n ), sum1, sum2],
       
       SimplifySum[c_ T[B___, A_, C___] T[D___, A_, E___], sum1___, {A_, __}, sum2___] :> SimplifySum[c/2 (T[B, E, D, C]  - T[B, C] T[D, E]/n ), sum1, sum2],
       SimplifySum[T[B___, A_, C___] T[D___, A_, E___], sum1___, {A_, __}, sum2___] :> SimplifySum[1/2 (T[B, E, D, C]  - T[B, C] T[D, E]/n ), sum1, sum2],
       SimplifySum[c_ T[B___, A_, C___]^2, sum1___, {A_, __}, sum2___] :> SimplifySum[c/2 (T[B, C, B, C]  - T[B, C]^2/n ), sum1, sum2],
       SimplifySum[T[B___, A_, C___]^2, sum1___, {A_, __}, sum2___] :> SimplifySum[1/2 (T[B, C, B, C]  - T[B, C]^2/n ), sum1, sum2]
    };

  \end{lstlisting}
  While the first three lines in \texttt{subSimplifySum} are the same as in \eq{advanced-relations}, the second and third block cover the last two lines in \eq{advanced-relations} respectively, and account for complications due to possible prefactors and squares in the expression. 
  Commencing with the evaluation, mass anomalous dimensions can be obtained in the usual way. 
    \begin{lstlisting}[escapeinside={|}{|}]
|\color{CodeGray}\textbf{In[1]:=}|  |{$\beta$}|[phi, phi, {1,1}, {1,1}, 0]
|\color{CodeGray}\textbf{Out[1]=}|  |\color{CodeOutput}m2/24 |
|\color{CodeGray}\textbf{In[2]:=}|  (4 |{$\mathtt{\pi}$}|)^2 24/m2 |{$\beta$}|[phi, phi, {1,1}, {1,1}, 1] // Expand
|\color{CodeGray}\textbf{Out[2]=}|  |\color{CodeOutput}2 $\lambda$1 + 2 n\verb!^!2 $\lambda$1 - 6 $\lambda$2/n + 4 n $\lambda$2|
|\color{CodeGray}\textbf{In[3]:=}|  (4 |{$\mathtt{\pi}$}|)^4 24/m2 |{$\beta$}|[phi, phi, {1,1}, {1,1}, 2] // Expand
|\color{CodeGray}\textbf{Out[3]=}|  |\color{CodeOutput}- 10 $\lambda$1\verb!^!2 - 10 n\verb!^!2 $\lambda$1\verb!^!2 + 60 $\lambda$1 $\lambda$2/n - 40 n $\lambda$1 $\lambda$2 + 30 $\lambda$2\verb!^!2 |
        |\color{CodeOutput} - 90 $\lambda$2\verb!^!2/n\verb!^!2  - 5 n\verb!^!2 $\lambda$2\verb!^!2|
  \end{lstlisting}
  In the quartic sector, the structures \texttt{T[\_\_\_]} will reappear at both tree and loop level, and $\beta$-functions can be extracted from their prefactors. Alternatively, the \texttt{T[\_\_\_]} may be completely removed by contracting external indices. In fact, we can even enforce the evaluation of the algebra as defined above by calling \texttt{SimplifyProduct[\_]} on the contraction defined in terms of \texttt{SimplifySum[\_\_]} instead of \texttt{Sum[\_\_]}.
   At tree level, one possibility is for instance 
  \begin{lstlisting}[escapeinside={|}{|}]
|\color{CodeGray}\textbf{In[4]:=}|  c1 = SimplifyProduct[
        SimplifySum[
         12 |{$\beta$}|[phi, phi, phi, phi, {a,1}, {a,1}, {1,1}, {1,1}, 0], 
         {a, 1, n^2-1}
        ]
      ]
|\color{CodeGray}\textbf{Out[4]=}|  |\color{CodeOutput}$\lambda$1 + n\verb!^!2 $\lambda$1 - 3 $\lambda$2/n + 2 n $\lambda$2|
|\color{CodeGray}\textbf{In[5]:=}|  c2 = SimplifyProduct[
           SimplifySum[
              96 T[a, b, c, d] |{$\beta$}|[phi, phi, phi, phi, {a,1}, {b,1}, {c,1}, {d,1}, 0], 
              {a, 1, n^2-1}, 
              {b, 1, n^2-1}, 
              {c, 1, n^2-1}, 
              {d, 1, n^2-1}
           ]
        ]
|\color{CodeGray}\textbf{Out[5]=}|  |\color{CodeOutput}6 $\lambda$1/n - 10 n $\lambda$1 + 4 n\verb!^!3 $\lambda$1 + 24 $\lambda$2 - 18 $\lambda$2/n\verb!^!2  - 7 n\verb!^!2 $\lambda$2 + n\verb!^!4 $\lambda$2 |
  \end{lstlisting}
  which is indeed free of the trace. This has to be matched up with the loop-level expressions
  \begin{lstlisting}[escapeinside={|}{|}]
|\color{CodeGray}\textbf{In[6]:=}|  b1 = SimplifyProduct[
           SimplifySum[
              12 (4 |$\pi$|)^2 |{$\beta$}|[phi, phi, phi, phi, {a,1}, {a,1}, {1,1}, {1,1}, 1], 
              {a, 1, n^2-1}
           ]
        ] // Simplify
|\color{CodeGray}\textbf{Out[6]=}|  |\color{CodeOutput}2 (7 + 8 n\verb!^!2 + n\verb!^!4) $\lambda$1\verb!^!2 + 4 (7 + n\verb!^!2) (-3 + 2 n\verb!^!2) $\lambda$1 $\lambda$2/n | 
        |\color{CodeOutput}+ 2 (63 - 30 n\verb!^!2 + 7 n\verb!^!4) $\lambda$2\verb!^!2/n\verb!^!2 |
|\color{CodeGray}\textbf{In[7]:=}|  b2 = SimplifyProduct[
           SimplifySum[
              96 (4 |$\pi$|)^2 T[a, b, c, d] |{$\beta$}|[phi, phi, phi, phi, {a,1}, {b,1}, {c,1}, {d,1}, 1], 
              {a, 1, n^2-1}, 
              {b, 1, n^2-1}, 
              {c, 1, n^2-1}, 
              {d, 1, n^2-1}
           ]
        ] // Simplify
|\color{CodeGray}\textbf{Out[7]=}|  |\color{CodeOutput}4 (21/n - 32 n +  9 n\verb!^!3 + 2 n\verb!^!5) $\lambda$1\verb!^!2 |
        |\color{CodeOutput}+ 8 (93 - 63/n\verb!^!2 - 37 n\verb!^!2 + 7 n\verb!^!4) $\lambda$1 $\lambda$2 |
        |\color{CodeOutput}+ 4 (-1 + n\verb!^!2) (-3 + n\verb!^!2)(63 - 6 n\verb!^!2  + n\verb!^!4) $\lambda$2\verb!^!2/n\verb!^!3|
  \end{lstlisting}
  which is left as an exercise to the reader.
  Putting everything together, we find that \ARGES provides us with the one and two loop anomalous dimensions and the one loop renormalisation group equations as
  \begin{equation}
  \begin{aligned}
    \gamma_{m^2}^{1\ell} &= 2 (n^2 + 1) \lambda_1 + \frac{4 n^2 - 6}{n}\,\lambda_2  , \\
    \gamma_{m^2}^{2\ell} &= - 10 (n^2 + 1) \lambda_1^2 - \frac{20 (2 n^2 - 3) }{n} \lambda_1 \lambda_2 - \frac{5(n^4 - 6 n^2 + 18 )}{n^2} \lambda_2^2,\\
    \beta_{\lambda_1}^{1\ell} &= 2 (7 +  n^2)  \lambda_1^2 + \frac{4(2 n^2 - 3)}{n}\, \lambda_1 \lambda_2   + \frac{6 (n^2 + 3)}{n^2} \lambda_2^2 , \\
    \beta_{\lambda_2}^{1\ell} &= 24 \lambda_1 \lambda_2 + \frac{4(n^2 - 9)}{n}\lambda_2^2\,.
  \end{aligned}
  \end{equation}
  This study can be extended to higher loop orders. At one-loop, our results confirm the findings of \cite{Hnatic:2020kyo} up to normalisation conventions.
  This concludes the illustration  of some of the more specialised capabilities of \ARGES.
  
 \subsection{Validity checks}
 The capabilities and the accuracy of  \ARGES have been tested extensively. This includes many cross-checks against  existing results in the literature, cross-checks against  the packages \SARAH and \pyrate, and cross-checks with the traditional ``by hand" extraction of expressions up to three loops, including for the model  \eqref{eq:LiSa-action} \cite{Bond:2017tbw}, and others.  \ARGES has  also been used as the primary tool for the derivation of $\beta$-functions in a number of studies  \cite{Bond:2017tbw,Schienbein:2018fsw,Bond:2019npq,Steudtner:2020tzo,Hiller:2020fbu,Steudtner:2020jcj}, 
 and we have confirmed that it represents a capable alternative to \SARAH and \pyrate for processing extensions of the Standard Model.

     \section{Summary \& conclusion}
The sheer increase of complexity with loop order or field content in the computation of renormalisation group equations in particle physics  necessitates powerful automation tools. 
Here, we have introduced \ARGES, a new framework for the computation of renormalisation group equations. 
A key design feature of \ARGES over other tools such as \SARAH and \pyrate  is that it follows a more algebraic approach to resolve index contractions.  
Most notably, \ARGES can handle models which no pre-existing framework has been able to process, including theories with general gauge groups and representations, unspecified multiplicity of matter fields, matrix scalars, highly complex  potentials, and even unknown vertex contractions.
Its capabilities cover any perturbatively renormalisable  quantum field theory with Lagrangian \eqref{eq:master-template}.
Setup, input and output of \ARGES are  straightforward, and its interactive API invites a quick and dynamic working style. 
At the same time, \ARGES is flexible and powerful, also offering a maximum of user control.  
We hope that \ARGES might prove useful  for theorists, model builders, and practitioners alike.

\paragraph{Acknowledgements}\\
We thank to Lohan Sartore and Alexander V.~Bednyakov for useful comments on the manuscript.
TS has been supported by the Deutsche Akademische Austauschdienst (DAAD) under the Grant 57314657. DL is supported by the Science and Technology Facilities Council (STFC) under the Consolidated Grant ST/T00102X/1.

\paragraph{Note added}\\ While finishing up this publication, we have been informed that a similar framework is currently in development \cite{RGBeta}, which is not associated with \ARGES. The project appears to be younger, but shares some of \ARGES features due to overlapping design goals. The most prominent difference is the use of the basis \cite{Poole:2019txl,Poole:2019kcm} for the template action. The corresponding preprint \cite{Thomsen:2021ncy} has appeared while this work was under review.

\bibliographystyle{jhep}
\bibliography{ref}
\end{document}